\documentclass{ar-1col}
\usepackage{amsmath}
\usepackage{amssymb}
\usepackage{amsthm}
\usepackage{amsfonts}
\usepackage{listings}
\usepackage{physics}
\usepackage{enumerate}
\usepackage{latexsym}
\usepackage{psfrag}
\usepackage{subcaption} 
\usepackage{bm}
\usepackage{graphicx}
\usepackage{blkarray}
\usepackage{array}
\usepackage{color}
\usepackage[normalem]{ulem}
\usepackage{hyperref}

\newcommand{\definition}[1]{\emph{#1}}
\usepackage[numbers]{natbib}

\def\beq{\begin{equation}}
\def\eeq{\end{equation}}
\def\bal{\begin{aligned}}
\def\eal{\end{aligned}}

\begin{document}

\title{Band Representations and Topological Quantum Chemistry}
\author{Jennifer Cano$^1{}^,{}^2$ and Barry Bradlyn$^3$
\affil{$^1$Department of Physics and Astronomy, Stony Brook University, Stony Brook, New York 11974, USA; email:jennifer.cano@stonybrook.edu}
\affil{$^2$Center for Computational Quantum Physics, The Flatiron Institute, New York, New York 10010, USA}
\affil{$^3$Department of Physics and Institute for Condensed Matter Theory, University of Illinois at Urbana-Champaign, Urbana, IL, 61801-3080, USA; email:bbradlyn@illinois.edu}}

\date{\today}
\begin{abstract}
    In this article, we provide a pedagogical review of the theory of topological quantum chemistry and topological crystalline insulators. We begin with an overview of the properties of crystal symmetry groups in position and momentum space. Next, we introduce the concept of a band representation, which quantifies the symmetry of topologically trivial band structures. By combining band representations with symmetry constraints on the connectivity of bands in momentum space, we show how topologically nontrivial bands can be catalogued and classified. We present several examples of new topological phases discovered using this paradigm, and conclude with an outlook towards future developments.
\end{abstract}


\maketitle

\tableofcontents

\section{Introduction}

One of the most transformative breakthroughs in the last few decades of condensed matter physics has been the discovery of topological phases of matter. Phenomena such as the integer and fractional quantum Hall effects \cite{klitzing1980new,laughlin1981quantized,tsui1982two,laughlin1983anomalous}, time-reversal invariant two- and three-dimensional topological insulators (TIs) \cite{Kane04,bernevig2006quantum,konig2007quantum,fukanemele,xia2009observation}, symmetry-protected topological band insulators \cite{Fu2011,Hsieh2012,Turner2010,Turner2012,Hourglass,Wieder17,hotis,benalcazar2017quantized,fangrotational,Aris2014,Fang2012,Hughes11}, and topological semimetals \cite{xu2015discovery,Lv15,Lv15a,Young2015,Steinberg14,Young12,Bradlyn2016,schroter2019chiral,Rao2019,Sanchez2018,Wan11,TangEA17} have revealed new surprises in ``complete'' topics such as the Landau theory of phase transitions and the band theory of solids. Topologically nontrivial materials exhibit robust transport properties such as the quantized Hall and magnetoelectric effects, edge states, and Fermi arcs. From a theoretical perspective, topological materials promise even more new developments in our understanding of physics.

Topology in systems of noninteracting electrons ultimately derives from the structure of Bloch states as a function of momentum. Because crystal symmetries relate Bloch states at different momenta (and sometimes even at the same momentum), they enrich the universe of protected topological phases beyond what it possible in the traditional Altland-Zirnbauer classes \cite{Fu2011,Shiozaki2014,Chiu2013,chiu2014,ando2015topological}. The recently developed theory of topological quantum chemistry (TQC) provides a theoretical and numerical recipe for understanding how topological crystalline bands arise from the interplay of localized atomic-like orbitals with the symmetries of a crystal \cite{NaturePaper,EBRTheoryPaper,GraphTheoryPaper}. Given a crystal symmetry group, topological quantum chemistry provides a map between the locations of atoms and orbitals within the crystal unit cell and allowed band structures. Wannier functions, which are the Fourier transform of the Bloch functions \cite{Marzari2012}, provide the link between the topology of Bloch functions in momentum space, and the localized orbital description of chemical compounds. In topologically trivial band structures, the Wannier functions are exponentially localized and respect the crystal symmetries. By contrast, topological crystalline bands are precisely those which do not admit a description in terms of exponentially localized, symmetric Wannier functions.

A set of bands arising from localized, symmetric Wannier functions form a representation of the crystal symmetry group known as a band representation \cite{Zak1980,Zak1981,Zak1982}, and all band representations can be built up from a finite collection of elementary band representations (EBRs) \cite{Michel2001}. The application of TQC has allowed for the discovery of new phenomena such as higher-order \cite{hotis} and fragile topological bands \cite{comment,Slager2018,bradlyn2019disconnected,Fragile2017}, and has enabled high-throughput searches for topological materials \cite{bigmaterials,bigmaterials-china,ashvin-materials}.

In this review, we will present a self-contained introduction to TQC. We will start in Section~\ref{sec:groups} with an introduction to the theory of crystal symmetry groups (space groups) and their representations, as they pertain to electrons moving in solids. Unlike the more familiar textbook treatments, we will emphasize how crystal symmetry constrains electrons in \emph{both} position and momentum space. To connect the position and momentum space pictures, we will in Sec.~\ref{sec:bandreps} construct the band representations of the space groups. Band representations give the fundamental building blocks of all electronic structures that can be connected to an atomic limit. In Sec.~\ref{sec:topology} we show how the theory of band representations can be used to define and distinguish topologically nontrivial band structures, focusing first on the ``symmetry indicated'' topological phases that differ from band representations at isolated points in momentum space. In Sec.~\ref{sec:beyond} we will show how TQC can be applied outside the paradigm of symmetry indicators. In Sec.~\ref{sec:fragile} we will introduce the concept of fragile topology, which emerged through the study of systems with a fixed number of occupied bands. Finally, in Sec.~\ref{sec:outlook} we will comment on future developments and applications. 


\section{Review of crystal symmetry}\label{sec:groups}

We briefly review point groups and space groups, assuming the reader is a physicist familiar with group theory. For a thorough introduction, we refer the reader to Ref~\cite{Cracknell}. Precise definitions of crystallographic concepts can be found in the International Tables for Crystallography \cite{ITA}.
The data described in this section 
can be found on the Bilbao Crystallographic Server (BCS) \cite{Bilbao1,Bilbao2,Bilbao3,Bilbao4}.

\subsection{Review of point and space groups}



A space group $G$ is generated by a subgroup of lattice translations, $T=\langle \mathbf{t}_i, i=1,2,3\rangle$, as well as a collection of other symmetry operations. 
Each symmetry $g\in G$ is denoted by:
\begin{equation}
g = \{ R| \mathbf{v} \},
\label{eq:symmetry}
\end{equation}
where $R$ is a point group operation (rotation, reflection, or identity) and $\mathbf{v}$ is a translation. $g$ acts on a spatial point $\mathbf{q}$ as
\begin{equation}
g\mathbf{q} = R\mathbf{q} + \mathbf{v}.
\end{equation}
We denote translations in 3D by:
\begin{equation}
\{ E | n_1\mathbf{t}_1 + n_2\mathbf{t}_2 + n_3\mathbf{t}_3 \} \equiv \{ E | n_1n_2n_3 \},
\end{equation}
where $E$ always denotes the identity point group operation.
Notice that a space group is an infinite group, since it includes an infinite number of translation elements.

A symmorphic space group is one which can be written as a semidirect product of a point group $P$ and the group of translations, i.e., $G = P \ltimes T$.
If $G$ is symmorphic, then for all elements $g\in G$, the translation $\mathbf{v}$ in Eq.~(\ref{eq:symmetry}) is a lattice translation, i.e., $\mathbf{v} = n_i\mathbf{t}_i$, where $n_i \in \mathbb{Z}$.
The remaining space groups are non-symmorphic, that is, for any choice of origin, there always exists $g\in G$ such that $\mathbf{v}$ is not a lattice translation.
Colloquially, glide and screw symmetries are sometimes referred to as non-symmorphic symmetries because a space group that contains a glide or screw symmetry must be a non-symmorphic space group.
However, this terminology is not precise because there are two non-symmorphic space groups that do not contain either glide or screw symmetries \cite{michel2001fundamental}.

The theory developed in this manuscript applies to crystallographic groups in any dimension.
The analogues of space groups in 2D are wallpaper groups.
Layer groups and rod groups describe the symmetry of 2D layers or 1D rods embedded in 3D space.

\subsection{Position space: Wyckoff positions and site-symmetry groups}
\label{sec:Wyckoff}

Let $G$ be a crystallographic group. 
For each point, or site, $\mathbf{q}$, in position space, the \definition{site-symmetry group}, or \definition{stabilizer group}, 
of $\mathbf{q}$, consists of the (finite) subgroup of $G$ that leaves $\mathbf{q}$ invariant, and is denoted 
\begin{equation}
G_\mathbf{q} \equiv \{ g | g\mathbf{q} = \mathbf{q} \} \subset G
\label{eq:defGq}
\end{equation}
While $G_\mathbf{q}$ may contain symmetry operations $\{ R | \mathbf{v} \}$ with non-zero translations (i.e., $\mathbf{v} \neq \mathbf{0}$),
by definition, $G_\mathbf{q}$ is always isomorphic to a crystallographic point group.

The set $\{ g\mathbf{q} | g\in G\} $ defines the orbit of a point $\mathbf{q}$.
It is straight-forward to show that the site-symmetry group of a point $\mathbf{q}'$ in the orbit of $\mathbf{q}$ is conjugate to $G_\mathbf{q}$ and therefore $G_{\mathbf{q}'}$ and $G_\mathbf{q}$ are isomorphic.
A \definition{Wyckoff position} is defined as a set of points whose site-symmetry groups are conjugate to each other; however, note that two points in the same Wyckoff position are not necessarily in the same orbit.
For example, in the space group $P2$ generated by $\{ C_{2z} | \mathbf{0} \}$ and lattice translations by $\hat{\mathbf{x}}$, $\hat{\mathbf{y}}$, and $\hat{\mathbf{z}}$, all points along the $\hat{\mathbf{z}}$ axis have the same site-symmetry group, generated by $\{ C_{2z} | \mathbf{0} \}$, and hence are all in the same Wyckoff position.


Given a particular point, $\mathbf{q}$, the \definition{multiplicity}, $n$, of the Wyckoff position containing $\mathbf{q}$ is given by the number of points in the orbit of $\mathbf{q}$ that reside in the conventional unit cell.
Each Wyckoff position is then given a label of the form $n\alpha$, where $n$ is the multiplicity of the Wyckoff position and $\alpha = a,b,c ,\dots$ is a letter that orders the Wyckoff positions in a particular space group by ascending $n$ (and serves to distinguish different Wyckoff positions with the same $n$).

\begin{figure}[h]
     \centering
     \begin{subfigure}{0.25\textwidth}
         \centering
         \includegraphics[width=\textwidth]{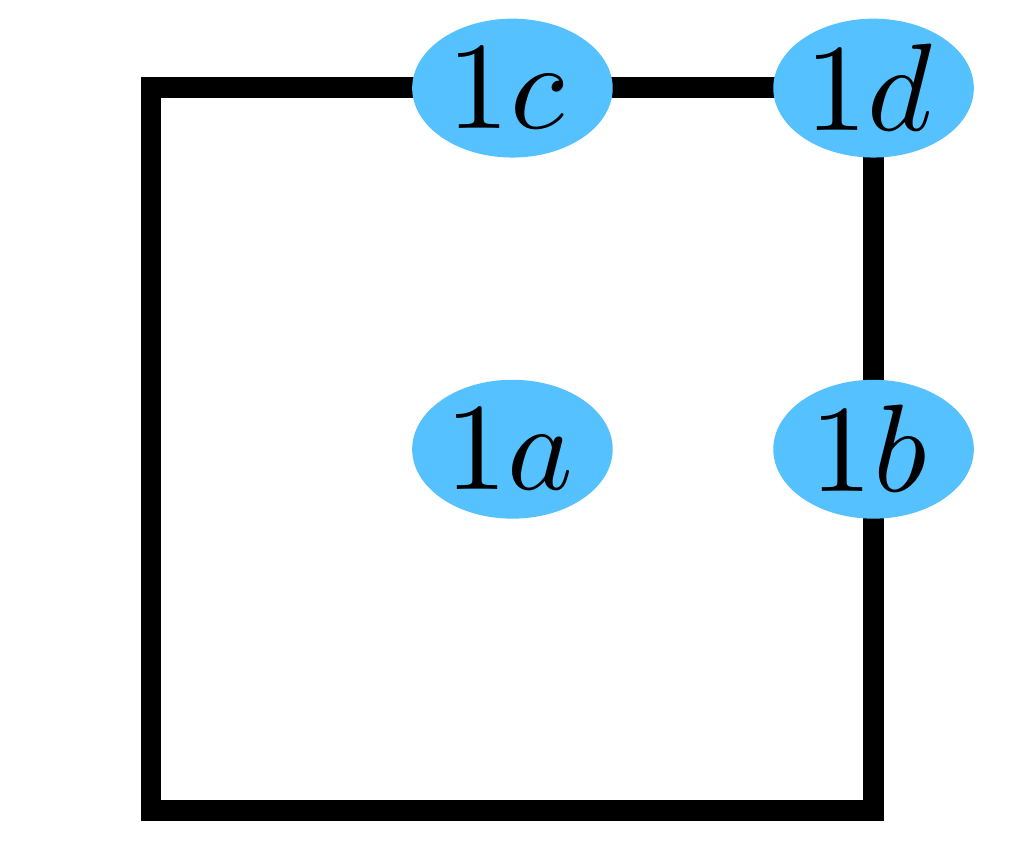}
         \vspace{-15pt}
         \caption{Real space}
         \label{fig:p2-Wyckoff}
     \end{subfigure}
     \quad 
     \begin{subfigure}{0.25\textwidth}
         \centering
         \includegraphics[width=\textwidth]{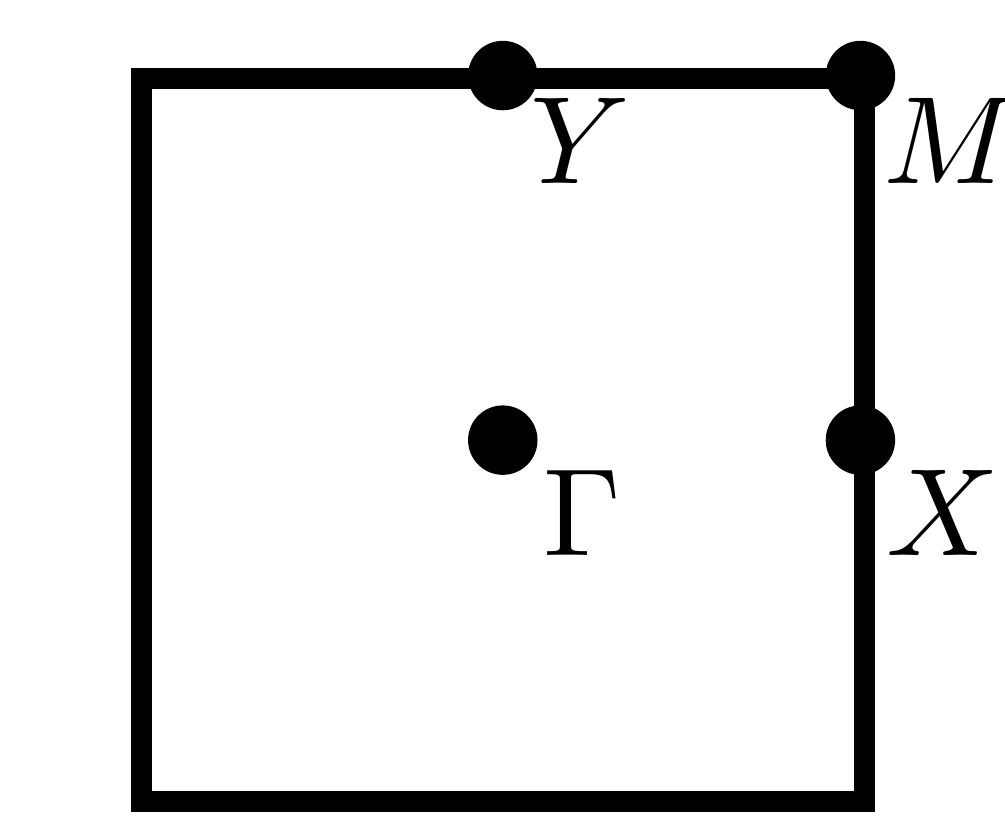}
         \vspace{-15pt}
         \caption{Momentum space}
         \label{fig:p2-BZ}
     \end{subfigure}
     \vspace{10pt}
        \caption{(a) Maximal Wyckoff positions and (b) high symmetry BZ points (b) in space group $P2$ or wallpaper group $p2$.}
        \label{fig:p2}
\end{figure}

For example, the space group $P2$ has five Wyckoff positions, $1a$, $1b$, $1c$, $1d$, and $2e$, shown in Fig.~\ref{fig:p2-Wyckoff}.
The $1a$ position contains the origin and all points along the $\hat{\mathbf{z}}$ axis; therefore for each point $\mathbf{q}_a$ in the $1a$ position, $G_{\mathbf{q}_a}$ is generated by $\{ C_{2z} | \mathbf{0} \}$ and is isomorphic to the point group $C_2$.
The $1b$ position contains points of the form $( \frac{1}{2},0,z)$; therefore for each point $\mathbf{q}_b$ in the $1b$ position, $G_{\mathbf{q}_b}$ is generated by $\{ C_{2z} | \hat{\mathbf{x}} \}$ and is also isomorphic to the point group $C_2$.
While $G_{\mathbf{q}_a}$ and $G_{\mathbf{q}_b}$ are isomorphic, they are not conjugate to each other; there is no symmetry $g\in G$ such that $g\mathbf{q}_a=\mathbf{q}_b$.
Similarly, the $1c$ position contains points of the form $(0,\frac{1}{2},z)$, whose site-symmetry group is generated by $\{ C_{2z} | \hat{\mathbf{y}} \}$ and the $1d$ position contains points of the form $(\frac{1}{2},\frac{1}{2},z)$, whose site-symmetry group is generated by $\{ C_{2z} | \hat{\mathbf{x} } + \hat{\mathbf{y}} \}$; the site-symmetry groups for points in the $1c$ and $1d$ positions are also isomorphic to $C_2$.
Finally, the $2e$ position contains pairs of points $\mathbf{q}_e = (x,y,z)$ and $\mathbf{q}_e' = (-x,-y,z)$, with site-symmetry groups $G_{\mathbf{q}_e} = G_{\mathbf{q}_e'} = \{ E | \mathbf{0} \}$.
The $2e$ position is called the general position because it contains points that are not invariant under any symmetries in the group.

\definition{Maximal Wyckoff positions} are those whose site-symmetry groups are not a proper subgroup of any other site-symmetry group.
For example, in $P2$, the $1a$, $1b$, $1c$, and $1d$ Wyckoff positions are maximal, while the $2e$ position is not maximal.
In Sec.~\ref{sec:EBR}, the elementary band representations will be labelled by maximal Wyckoff positions.

\subsection{Momentum space: k-stars and little groups}
\label{sec:littlegroup}

The translation generators of a space group determine its Bravais lattice, which determines its Brillouin zone (BZ).
We will use $\mathbf{g}_i$ to denote a set of reciprocal lattice vectors satisfying $\mathbf{g}_i \cdot \mathbf{t}_j = 2\pi \delta_{ij}$.
Points in the BZ are denoted by $\mathbf{k}$.
The action of $g=\{R|\mathbf{v}\}\in G$ on $\mathbf{k}$ is $g\mathbf{k} = R\mathbf{k}$; colloquially, translations do not act in momentum space.

The \definition{little group} of a point $\mathbf{k}$, denoted $G_\mathbf{k}$, consists of the set of space group symmetries that leave $\mathbf{k}$ invariant up to a reciprocal lattice vector, i.e., $G_\mathbf{k} \equiv \{ g | g\mathbf{k} = \mathbf{k} + n_i \mathbf{g}_i \} \subset G$.
Notice that $G_\mathbf{k}$ is always infinite because it contains all lattice translations.
In the same vein, notice that glide, screw, and translation symmetries do not leave any points in position space invariant and therefore are not in any site-symmetry groups, but can be contained in the little groups.
The \definition{little co-group} of $\mathbf{k}$, denoted $\tilde{G}_\mathbf{k}$, is the finite group defined by modding out $G_\mathbf{k}$ by the subgroup $T\subset G$ of lattice translations, i.e., $\tilde{G}_\mathbf{k} = G_\mathbf{k} / T$.
The little co-group is also isomorphic to a point group; the little co-group in momentum space is analogous to the site-symmetry group in position space.

As an example, consider the wallpaper group $p2$, which has the same generators as $P2$ except $t_\mathbf{z}$.
In $p2$, there are four high-symmetry points in the BZ: $\Gamma=(0,0)$, $X=(\pi,0)$, $Y=(0,\pi)$, and $M=(\pi,\pi)$, shown in Fig.~\ref{fig:p2-BZ}.
Since each high-symmetry point is invariant (modulo a reciprocal lattice vector) under all space group operations, their little groups are equal to the full space group, i.e., $G_\Gamma = G_X = G_Y = G_M = G$.
(The little group of $\Gamma$ is always equal to the full space group.)
The little group of any other point is $T$, the group of translations.
Therefore, the little co-group of each high-symmetry point is isomorphic to the point group $C_2$, while the little co-group of a non-high-symmetry point is trivial.

The analog of a Wyckoff position in position space is a star in momentum space: the \definition{star} of a point $\mathbf{k}$, denoted $\mathbf{k}^\star$, consists of all points $\{g\mathbf{k}|g\in G\}$ in the BZ.
The little group of a point $\mathbf{k'}\in\mathbf{k}^\star$ is conjugate to $G_\mathbf{k}$. Irreps of the space groups are labeled by $\mathbf{k}^\star$, and are induced (see Sec.~\ref{sec:induction}) from little group irreps \cite{Cracknell}.

\section{Band Representations}\label{sec:bandreps}


Zak introduced band representations \cite{Zak1980,Zak1981,Zak1982} to describe the symmetry of an entire band in a band structure.
Traditionally, the textbook approach \cite{Kittel87,ashcroft2005solid} towards analyzing a band structure is to determine the symmetry at a single $\mathbf{k}$, where Bloch wave functions form the basis of the representations of the little group and their energetics can be described perturbatively by a $\mathbf{k}\cdot\mathbf{p}$ theory.
In contrast, the basis of a band representation is a set of symmetry-adapted Wannier functions \cite{Marzari2012}, localized in position space, whose energetics are described by a tight-binding model.

A band representation can be decomposed into a direct sum of representations at each $\mathbf{k}^\star$, but carries additional information about how the representations at different $\mathbf{k}^\star$s are related to each other, giving rise, for example, to the Berry-Zak phase and accompanying polarization \cite{Zakphase,ksv}. 
The appearance of the Zak phase hints at the connection between band representations and topology \cite{NaturePaper}.


We now derive the group theory of band representations.
In Sec.~\ref{sec:topology}, we describe the connection between band representations and topological band theory.

\subsection{Induced representations}
\label{sec:induction}


We first present a general construction to build an induced representation of $G$ from a representation of a subgroup $H$ (for a review of representation theory, see, for example, the book by Serre \cite{Serre}).
We will then apply this construction to band representations.

Given a group, $G$, a subgroup, $H$, and a coset decomposition of $H$ in $G$, 
\begin{equation}
    G = \bigcup_\alpha g_\alpha H \label{eq:defcoset}
\end{equation}
each representation $\rho$ of $H$ generates an \definition{induced representation} of $G$, which we denote $\rho_G \equiv \rho \uparrow G$.

A representation $\rho_G$ can be explicitly constructed from the representation $\rho$.
Specifically, if the rows/columns of $\rho$ are indexed by $i,j$, then the rows/columns of $\rho_G$ are indexed by $i\alpha, j\beta$, where $\alpha, \beta$ run over the cosets $g_\alpha H$ in Eq.~(\ref{eq:defcoset}).
The representation $\rho_G$ can then be written:
\begin{equation}
    \left[ \rho_G(h) \right]_{i\alpha,j\beta} = \left[ \tilde{\rho}(g_\alpha^{-1} h g_\beta) \right]_{ij},
    \label{eq:induction}
\end{equation}
where $h\in G$ and
\begin{equation}
    \left[ \tilde{\rho}(g ) \right]_{ij} = \begin{cases} \left[ \rho(g) \right]_{ij} & \text{if } g \in H \\ 0 & \text{else} \end{cases}
    \label{eq:defrhotilde}
\end{equation}

\subsection{Band representation construction via induction in position space}
\label{sec:bandinduction}

We now apply the construction in the previous section to build a representation of a space group induced from a representation of a site-symmetry group.
Let $G$ denote a space group and let $G_\mathbf{q}$ be the site-symmmetry group of some site, $\mathbf{q}$.
We seek a coset decomposition of $G_\mathbf{q}$ in $G$.
To this end, define the set $\lbrace \mathbf{q}_\alpha \rbrace$, $\alpha =1, 2, \dots, n$, to be the sites in the Wyckoff position of $\mathbf{q}$ residing in the primitive unit cell, defining $\mathbf{q}_1\equiv \mathbf{q}$.
Then for each $\mathbf{q}_\alpha$, choose a space group element $g_\alpha \in G$ such that $\mathbf{q}_\alpha = g_\alpha \mathbf{q}$.
(Different choices of unit cell and of $g_\alpha$ change the basis for the induced representation.)
The $g_\alpha$, combined with translations, $T$, generate a decomposition of $G$ with respect to $G_\mathbf{q}$:
\begin{equation}
    G = \bigcup_{\alpha=1}^n g_\alpha \left( G_\mathbf{q}\ltimes T\right)
    \label{eq:coset}
\end{equation}

Given a representation $\rho$ of $G_\mathbf{q}$, we can construct the induced representation of $G$ by generalizing Eq.~(\ref{eq:induction}); specifically:
\begin{equation}
    \left[ \rho_G(h) \right]_{(i,\alpha,\mathbf{t}),(j,\beta,\mathbf{t}')} = \left[ \tilde{\rho}\left(g_\alpha^{-1} \{ E | \mathbf{t} \} h \{ E | \mathbf{t}' \}^{-1} g_\beta \right) \right]_{ij},
    \label{eq:rhoG}
\end{equation}
where $\tilde{\rho}$ is defined by Eq.~(\ref{eq:defrhotilde}).
Notice that $\rho_G(h)$ is a representation of the entire space group, i.e., Eq.~(\ref{eq:rhoG}) explicitly gives a matrix for each symmetry $h\in G$.

A \definition{band representation} of a space group, $G$, is a direct sum of representations, each induced from a representation of the site-symmetry group of a Wyckoff position in $G$.
Denoting the Wyckoff positions in $G$ by $n\alpha$ (defined in Sec.~\ref{sec:Wyckoff}), 
and letting $\mathbf{q}_\alpha$ label a representative site in each Wyckoff position, then the most general band representation of $G$ takes the form  
\begin{equation}
    \bigoplus_\alpha \left( \rho_\alpha \uparrow G \right),
    \label{eq:BRsum}
\end{equation}
where $\rho_\alpha$ is a representation of the site-symmetry group $G_{\mathbf{q}_\alpha}$.



\subsection{Wannier basis}
\label{sec:Wannier}

To unpack the definition of the induced representation in Eq.~(\ref{eq:rhoG}), it is helpful to define a basis for $\rho_G$, which will turn out to be a set of Wannier functions.

Let $W_{i1}(\mathbf{r})$, $i=1,\dots {\rm dim}(\rho)$, be a set of (Wannier) functions localized on $\mathbf{q}$ that transform under the representation $\rho$ of $G_\mathbf{q}$ such that for each $g\in G_\mathbf{q}$:
\begin{equation}
g W_{i1}(\mathbf{r}) = \left[ \rho(g) \right]_{ji} W_{j1}(\mathbf{r})
\label{eq:Wannierg}
\end{equation}

Within the primitive unit cell, a Wannier function localized on each $\mathbf{q}_\alpha$ can be defined as:
\begin{equation}
W_{i\alpha}(\mathbf{r}) = g_\alpha W_{i1}(\mathbf{r}) = W_{i1}(g_\alpha^{-1}\mathbf{r})
\label{eq:Wanniergalpha}
\end{equation}
By extension, translated counterparts in other unit cells are defined by:
\begin{equation}
\lbrace E | \mathbf{t} \rbrace W_{i\alpha}(\mathbf{r}) = W_{i\alpha} (\mathbf{r} - \mathbf{t}),
\label{eq:Wanniertrans}
\end{equation}
where $\mathbf{t}$ is a lattice vector.
The set of $n\times {\rm dim}(\rho) \times \mathcal{N}$ functions $W_{i\alpha}(\mathbf{r} - \mathbf{t})$, where $\mathcal{N} \rightarrow \infty$ is the number of unit cells in the system, are exactly the basis states on which the induced representation $\rho_G$ acts.
Specifically, given $h=\{ R | \mathbf{v} \}\in G$, the coset decomposition (\ref{eq:coset}) implies that for each $g_\alpha$, the combined operation $hg_\alpha$ can be decomposed as:
\begin{equation}
hg_\alpha = \{ E | \mathbf{t}_{\beta\alpha}\} g_\beta g,
\label{eq:hgalphacoset}
\end{equation}
for a unique choice of coset $g_\beta H$, $g\in G_\mathbf{q}$, and lattice vector $\mathbf{t}_{\beta\alpha} \equiv h\mathbf{q}_\alpha - \mathbf{q}_\beta$.
Combining the decomposition in Eq.~(\ref{eq:hgalphacoset}) with the action of $g\in G_\mathbf{q}$ in Eq.~(\ref{eq:Wannierg}) and the definitions of the transformed Wannier functions in Eqs.~(\ref{eq:Wanniergalpha}) and (\ref{eq:Wanniertrans}), we see that the Wannier functions transform in the induced representation $\rho_G$, according to:
\begin{equation}
\rho_G(h) W_{i\alpha}(\mathbf{r} - \mathbf{t}) = \sum_{j=1}^{\dim(\rho)} \left[ \rho(g) \right]_{ji} W_{j\beta} (\mathbf{r} - R\mathbf{t} - \mathbf{t}_{\beta\alpha} ),
\label{eq:rhoG2}
\end{equation}
where we sum over $j$ on the right-hand-side, but $\beta, g$ and $\mathbf{t}_{\beta\alpha}$ are uniquely determined by the coset decomposition in Eq.~(\ref{eq:hgalphacoset}).
(The explicit derivation of Eq.~(\ref{eq:rhoG2}) is given in Eq.~(B1) of Ref.~\cite{EBRTheoryPaper}.)
In summary, the matrix representation of $\rho_G$ defined in Eq.~(\ref{eq:rhoG}) is written in the basis of the Wannier functions defined in Eqs.~(\ref{eq:Wannierg}), (\ref{eq:Wanniergalpha}) and (\ref{eq:Wanniertrans}).

\subsection{Band representations in momentum space}
\label{sec:bandrepmomentum}

While it is natural to build a band representation in position space, 
it will be useful to view a band representation in momentum space. 
To this end, we define the Fourier transformed Wannier functions:
\begin{equation}
    a_{i\alpha}(\mathbf{k},\mathbf{r}) = \sum_\mathbf{t} e^{i\mathbf{k} \cdot \mathbf{t}} W_{i\alpha}(\mathbf{r} - \mathbf{t}),
\end{equation}
where the sum is over all lattice vectors, $\mathbf{t}\in T$.
The Fourier transform amounts to a unitary transformation that exchanges $\mathcal{N}$ unit cells in the system for $\mathcal{N}$ distinct $\mathbf{k}$ points.
The action of $\rho_G$ in momentum space becomes \cite{EBRTheoryPaper}: 
\begin{equation}
    \rho_G(h) a_{i\alpha}(\mathbf{k},\mathbf{r}) = e^{-i(R\mathbf{k})\cdot\mathbf{t}_{\beta\alpha}} \sum_{j=1}^{\dim(\rho)} \left[ \rho(g)\right]_{ji} a_{j\beta}(R\mathbf{k},\mathbf{r}),
    \label{eq:rhoG3}
\end{equation}
where, as in Eq.~(\ref{eq:rhoG2}), $\beta, g$ and $\mathbf{t}_{\beta\alpha}$ are uniquely determined by the coset decomposition in Eq.~(\ref{eq:hgalphacoset}).

In momentum space, the matrix representation of $\rho_G$ can be interpreted as an $\mathcal{N} \times \mathcal{N}$ matrix of $n\dim (\rho) \times n\dim(\rho)$ blocks (recall $n$ is the number of coset representatives $g_\alpha$), where each block is labelled by $\mathbf{k},\mathbf{k}'$.
Most of the blocks are zero:
given $h=\{ R|\mathbf{v} \} \in G$, there is only one non-zero block in each row and column, corresponding to $\mathbf{k}'=R\mathbf{k}$.
We denote this block $\rho_G^\mathbf{k}(h)$.
Notice that the band representation is completely defined by the set of nonzero blocks $\rho_G^\mathbf{k}(h)$, for all $\mathbf{k}$ in the first BZ and all $h\in G$.

\subsection{Little group representations from band representations}
\label{sec:subduction}

A non-zero block $\rho_G^\mathbf{k}(h)$ will be a diagonal block in $\mathbf{k}$ if and only if $h\mathbf{k} = \mathbf{k}$ up to a reciprocal lattice vector, i.e., exactly when $h \in G_\mathbf{k}$, where $G_\mathbf{k}$ is the little group of $\mathbf{k}$. 
For a given $\mathbf{k}$, the set of $\rho_G^\mathbf{k}(h)$, where $h\in G_\mathbf{k}$, form a representation of $G_\mathbf{k}$ that we denote $\rho_G\downarrow G_\mathbf{k}$. 

Consequently, a band representation can be labelled by the set of representations of the little group, $\rho_G\downarrow G_\mathbf{k}$, at each $\mathbf{k}$.
This labeling misses information about how the representations at different $\mathbf{k}$ are connected, and hence does not distinguish between all pairs of band representations.
Nonetheless, it is a valuable tool to diagnose topological phases, as we detail in Sec.~\ref{sec:symmetryindicators}.

\subsection{Example: band representations in $p2$}
\label{sec:p2BR}

As an example, we return to the wallpaper group $p2$.
Recall from Secs.~\ref{sec:Wyckoff} and \ref{sec:littlegroup}, there are four maximal Wyckoff positions, $1a$, $1b$, $1c$, and $1d$, shown in Fig.~\ref{fig:p2-Wyckoff}, each of whose site-symmetry group is isomorphic to the point group $C_2$. 
The point group $C_2$ has two irreps, $A$ and $B$, as shown in Table~\ref{tab:c2chars}.
The $A/B$ irrep of the site-symmetry group $G_{1\alpha}$, where $\alpha=a,b,c,d$, induces a band representation denoted $\rho_{G,1\alpha,A/B}$.
We derive the matrix form of $\rho_{G,1\alpha,A/B}$ by applying Eq.~(\ref{eq:rhoG3}) to each $h$.
Omitting the indices $i,j$ because each irrep is one-dimensional and the indices $\alpha,\beta$ because each maximal Wyckoff position has multiplicity one,
we arrive at the band representations:
\begin{equation}
    \rho_{G,1\alpha,A/B} (h)= \begin{cases}
        e^{-i\mathbf{k} \cdot \mathbf{t}} & h=\{ E | \mathbf{t} \} \\
        \pm e^{i\mathbf{k} \cdot (\mathbf{t} -2\mathbf{q}_\alpha)} &h=\{ C_2 | \mathbf{t} \}
    \end{cases}
    \label{eq:p2BR}
\end{equation}
The first line of Eq.~(\ref{eq:p2BR}) shows that translations are represented by a Bloch phase in all band representations.
The second line shows that the band representations $\rho_{G,1\alpha,A/B}$ can be distinguished by their $\{ C_2 | \mathbf{0} \}$ eigenvalues (i.e., little group irreps) at high-symmetry points in the BZ, ($\Gamma$, $X$, $Y$, $M$, shown in Fig.~\ref{fig:p2-Wyckoff}), which are indicated in Table~\ref{tab:p2bandreps}.

\begin{table}[t]
\begin{center}
\begin{tabular}{c|cc}
$\rho$ & $\chi(E)$ & $\chi(C_2)$ \\
\hline
A & 1 & 1 \\
B & 1 & -1
\end{tabular}
\end{center}
\caption{Character table for the point group $C_2$. Representations are distinguished by their character of the rotation $C_2$, positive for A and negative for B.}\label{tab:c2chars}
\end{table}

\begin{table}[t]
\begin{center}
\begin{tabular}{c|cccc}
$\rho$ & $\Gamma$ & $X$ & $Y$ & $M$\\
\hline
$1a, A$ & $+$ & $+$ & $+$ & $+$ \\
$1a, B$ & $-$ & $-$ & $-$ & $-$ \\
$1b, A$ & $+$ & $-$ & $+$ & $-$\\
$1b, B$ & $-$ & $+$ & $-$ & $+$\\
$1c, A$ & $+$ & $+$ & $-$ & $-$\\
$1c, B$ & $-$ & $-$ & $+$ & $+$\\
$1d, A$ & $+$ & $-$ & $-$ & $+$\\
$1d, B$ & $-$ & $+$ & $+$ & $-$
\end{tabular}
\end{center}
\caption{$C_2$ eigenvalues in the eight elementary band representations of $p2$. Each band representation has either $0$, $2,$ or $4$ negative eigenvalues.}\label{tab:p2bandreps}
\end{table}

\subsection{Composite and elementary band representations}
\label{sec:EBR}


We now define a notion of equivalence between two band representations, following Ref.~\cite{NaturePaper}:
two band representations $\rho_G$ and $\sigma_G$ are \definition{equivalent} if and only if there exists a unitary matrix-valued function $S(\mathbf{k},t,g)$ smooth in $\mathbf{k}$ and continuous in $t$ such that for all $g\in G$, $S(\mathbf{k},0,g) = \rho_G^\mathbf{k}(g)$, $S(\mathbf{k},1,g) = \sigma_G^\mathbf{k}(g)$ and $S(\mathbf{k},t,g)$ defines a band representation for $t\in[0,1]$.
This definition of equivalence is stronger than requiring the same little group representations at each $\mathbf{k}$:
it also implies that equivalent band representations share all Berry phases and Wilson loop \cite{ArisCohomology} invariants.
This distinction is nontrivial: it is possible for band representations to share the same little group representations at each $\mathbf{k}$ but differ by a Berry phase \cite{Bacry1988b,Michel1992,Bacry1993,hiddenpaper}.

We can thus define a \definition{composite} band representation as a band representation that is equivalent to a direct sum of two or more other band representations.
An \definition{elementary} band representation (EBR) is a band representation that is not composite. 

Notice that EBRs are not irreps of the space group because they can be block diagonalized into blocks corresponding to each $\mathbf{k}^\star$. 
Instead, EBRs are a minimal basis for band representations.
There are a finite number of EBRs \cite{NaturePaper}, indexed by irreps of maximal Wyckoff positions. They are enumerated on the BCS \cite{progbandrep,GroupTheoryPaper}.

Spin-orbit coupling is automatically incorporated into the formalism by using the double-valued $(SU(2))$ representations of the site-symmetry groups.
Similarly, time-reversal symmetric band representations are induced from time-reversal symmetric irreps of the site-symmetry groups.

\section{Topological systems are not Band representations}
\label{sec:topology}
Every band representation corresponds to a band structure with exponentially localized Wannier functions. 
Band representations, then, describe sets of bands that are adiabatically connected to an \definition{atomic limit}, as defined in Ref.~\cite{EBRTheoryPaper}.
It follows that topological bands are not band representations.
We now  derive constraints for bands that are not band representations and use them to classify topological bands.

\subsection{Compatibility relations and quasi-band reps}\label{sec:compatibility}

From a purely momentum-space perspective, we can define a band structure by specifying a set of little group representations at every point in the BZ, subject to consistency conditions. 
The simplest consistency condition is that the dimension $N$ of the little group representation at every $\mathbf{k}$ be the same; this is the statement that the band structure has $N$ bands. 
Less trivial constraints can be derived by relating the little groups of neighboring points in the BZ.
Consider a representation $\rho$ of the little group $G_{\mathbf{k}_0}$ of some high-symmetry point $\mathbf{k}_0$. At a neighboring point $\mathbf{k}_1=\mathbf{k}_0+d\mathbf{k}$, continuity of the Brillouin zone necessitates that
\begin{equation}
    G_{\mathbf{k}_1}\subseteq G_{\mathbf{k}_0},\label{eq:compsubgrp}
\end{equation}
i.e. each high-symmetry $\mathbf{k}$-point is also a member of a lower symmetry $\mathbf{k}$-manifold (line, plane, etc.). For instance, if $\mathbf{k}_0=\Gamma=(0,0,0)$, then $\mathbf{k}_1$ may be a point on the high-symmetry line $\Lambda=(x,x,x)\ni\Gamma$. It follows that the basis for the representation $\rho$ of $G_{\mathbf{k}_0}$ must transform under the 
subduced representation
\begin{equation}
    \rho_1=\rho\downarrow G_{\mathbf{k}_1}\label{eq:compatibility}
\end{equation}
of $G_{\mathbf{k}_1}$, formed by ``forgetting" those elements in $G_{\mathbf{k}_0}$ that are not in $G_{\mathbf{k}_1}$. 

Conditions such as Eq.~(\ref{eq:compatibility}) are known as \emph{compatibility relations}. For a collection of little group representations to form a band structure, the compatibility relations must be satisfied at every $\mathbf{k}$-point: for every pair $\mathbf{k}_0$ and $\mathbf{k}_1=\mathbf{k}_0+d\mathbf{k}$ of connected $\mathbf{k}$ points, if the representation $\rho$ of $G_{\mathbf{k}_0}$ appears in the band structure, then the representation $\rho\downarrow G_{\mathbf{k}_1}$ must occur at the point $\mathbf{k}_1$. Following Ref.~\cite{Bacry1993}, we refer to any set of little group representations satisfying these compatibility relations as a quasi-band representation.
The compatibility relations for all 230 space groups can be found on the BCS for both single and double-valued representations \cite{GroupTheoryPaper}.

It is important to note that every band representation is a quasi-band representation. The induction procedure outlined in Sec.~\ref{sec:bandinduction} guarantees that the compatibility relations are satisfied between every pair of connected $\mathbf{k}$-points, since the blocks $\rho_\mathbf{k}$ are continuous functions of $\mathbf{k}$. However, quasi-band representations exist which are not themselves band representations: {they preserve all crystal symmetries in momentum space, but lack exponentially localized Wannier functions. Such quasi-band representations are exactly topological bands.}

\begin{figure*}
\includegraphics[width=0.95\textwidth]{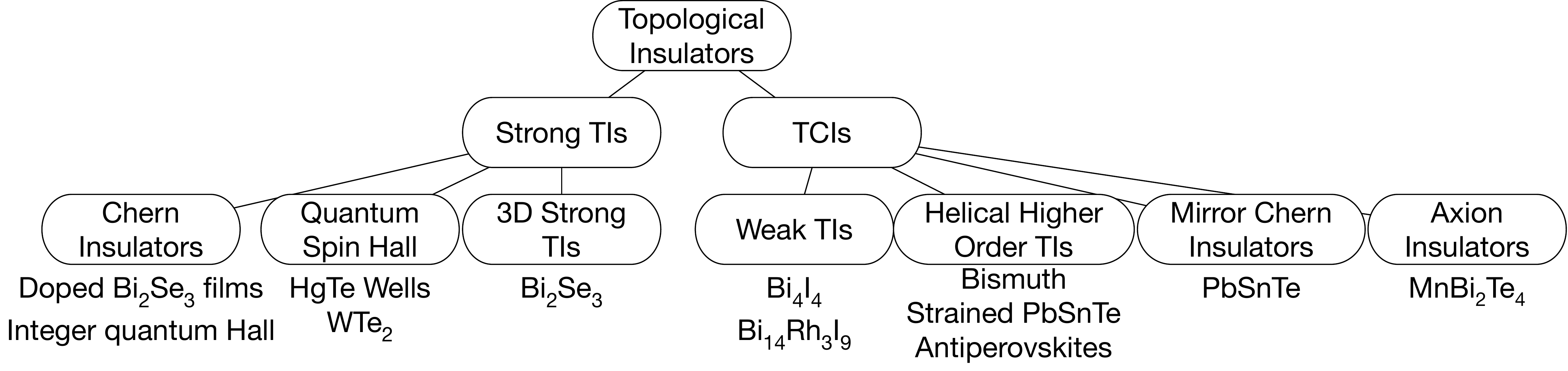}
\caption{Organizational chart outlining the different classes of free fermion topological phases, with material examples of each.}\label{fig:schematic}
\end{figure*}

{ In the remainder of this section, we will show how TQC can be used to distinguish trivial and topological bands. To orient ourselves,} we give in Fig.~\ref{fig:schematic} an organizational chart of the different topological free fermion phases. We restrict ourselves to the physically relevant case of $d\leq 3$ dimensions, and focus on fermionic systems with half-integer spin. 
Broadly, the two main classes of (stable) TIs are ``Strong TIs" and ``Topological crystalline insulators (TCIs).'' Strong TIs are robust to perturbations that break all crystal symmetries {and are classified by the famous ``Tenfold way'' \cite{ryu2010topological,schnyder2008classification,KitaevClassify}.} In two dimensions these can be Chern insulators (which require no symmetry) or quantum spin Hall insulators (which require time-reversal symmetry). In three dimensions, there are strong $\mathbb{Z}_2$ TIs with time-reversal symmetry. The realm of TCIs, on the other hand, is much richer. TCIs include weak TIs \cite{fukanemele,balents2007topological,roy2009topological}, protected by translation symmetry, which are adiabatically deformable to stacks of strong TIs in lower dimensions. Additionally, there are mirror Chern insulators \cite{Hsieh2012}, which feature an even number of surface Dirac cones on mirror-symmetric surfaces, as well as chiral (axion insulators) and helical higher-order TIs \cite{hotis,wieder2018axion,Po2017,witten2016fermion,Turner2010,Hughes11}, which host subdimensional hinge and corner states. 

We now describe when topology can be inferred directly from the little group representations in a quasi-band representation. 
We will present examples of both strong TIs and TCIs that can be described in this way.
In Sec.~\ref{sec:beyond} we will examine topological bands that cannot be diagnosed by symmetry eigenvalues alone, and show how TQC sheds light on these cases.

\subsection{Symmetry indicated phases: Smith normal form}
\label{sec:symmetryindicators}

When topological bands can be distinguished by their little group representations, we refer to them as ``symmetry indicated'' topological bands. Following Ref.~\cite{Po2017}, we formalize this notion by mapping band structures in each space group to a vector space $V_G$ (see also Ref.~\cite{po2020symmetry} for an alternative perspective). The dimension of this vector space is equal to the number of irreducible representations of the little groups of all symmetry-inequivalent classes $[\mathbf{k}]$ of points in the BZ. A natural basis for this vector space is given by the irreps $\rho_{i[\mathbf{k}]}$, where $i$ indexes the irreps of the little group in each $[\mathbf{k}]$ class. A vector
\begin{equation}
    \mathbf{v}=\sum_{i,[\mathbf{k}]} n_{i[\mathbf{k}]}\rho_{i[\mathbf{k}]}\label{eq:generalvector}
\end{equation}
with non-negative integer entries $\{n_{i[\mathbf{k}]}\}$ gives an assignment of little group irreps to points in the Brillouin zone. Every quasi-band representation maps to a vector in which the $\{n_{i[\mathbf{k}]}\}$ satisfy the compatibility relations of the space group. Similarly, each elementary band representation $\rho_\mathbf{k}^a$ maps to a vector $\mathbf{e}_a$ in the vector space, such that every atomic limit band structure can be identified with a vector 
\begin{equation}
    \mathbf{a}=\sum_{a} n_a\mathbf{e}_a
\end{equation}
with non-negative coefficients $n_a$.



As an example, let us return to the wallpaper group $p2$. There are four classes of $\mathbf{k}$ points with little co-group isomorphic to the point group $C_2$ ($\Gamma,X,Y$ and $M$), each with two one-dimensional irreducible representations. All other $\mathbf{k}$ points fall into the general position $GP$; their little groups have only one one-dimensional irreducible representation.
Using the notation of Tables~\ref{tab:c2chars} and \ref{tab:p2bandreps}, we write the nine basis elements for $V_G$ as $\Gamma_+,\Gamma_-, X_+,X_-, Y_+, Y_-,M_+,M_-,$ and $GP$, where the subscript indicates the $C_2$ eigenvalue. 
Since there are no high-symmetry lines in this wallpaper group, the only compatibility relation is that the dimension of the representations at each $\mathbf{k}$-point be the same. Thus, a vector corresponds to a quasi-band representation if and only if it satisfies
\begin{equation}
    n_{\Gamma+}+n_{\Gamma-}=n_{X+}+n_{X-}=n_{Y+}+n_{Y-}=n_{M+}+n_{M-}=n_{GP}
\end{equation}
with all $n_{i[\mathbf{k}]}$ non-negative. Writing out the coefficients as a vector in this basis, i.e. as $(n_{\Gamma+},n_{\Gamma-},n_{X+},n_{X-},n_{Y+},n_{Y-},n_{M+},n_{M-},n_{GP})$, we find that every quasi-band representation has the form
\begin{equation}
    \mathbf{v}=(n_{\Gamma+},N-n_{\Gamma+},n_{X+},N-n_{X+},n_{Y+},N-n_{Y+},n_{M+},N-n_{M+},N).\label{eq:qbrp2}
\end{equation}
Similarly, each elementary band representation from Table~\ref{tab:p2bandreps} maps to a vector $\mathbf{e}_a$, given by:
\begin{align}
    \mathbf{e}_{1aA}&=(1,0,1,0,1,0,1,0,1)^T,\qquad
    \mathbf{e}_{1aB}=(0,1,0,1,0,1,0,1,1)^T, \label{eq:ebrvec1}\\
    \mathbf{e}_{1bA}&=(1,0,0,1,1,0,0,1,1)^T ,\qquad
    \mathbf{e}_{1bB}=(0,1,1,0,0,1,1,0,1)^T, \\
    \mathbf{e}_{1cA}&=(1,0,1,0,0,1,0,1,1)^T,\qquad
    \mathbf{e}_{1cB}=(0,1,0,1,1,0,1,0,1)^T, \\
    \mathbf{e}_{1dA}&=(1,0,0,1,0,1,1,0,1)^T,\qquad
    \mathbf{e}_{1dB}=(0,1,1,0,1,0,0,1,1)^T\label{eq:ebrvec2},
\end{align}
where $N=1$ for a single band.

The question of identifying and classifying symmetry-indicated topological bands amounts to finding non-negative integer vectors $\mathbf{v}$ which are not expressible as non-negative integer sums of EBR vectors $\mathbf{e}_a$. Note that there are two ways that a quasi-band representation $\mathbf{v}$ can fail to be a symmetry-indicated atomic-limit band structure. First, it could be that $\mathbf{v}$ cannot be expressed as any \emph{integer} linear combination of EBR vectors, without regard to positivity or negativity of coefficients. We will refer to such quasi-band reps as \emph{stable} symmetry indicated topological bands. These are so named because a stable symmetry indicated topological band remains symmetry indicated under addition of trivial (occupied) bands. By contrast, we define \emph{fragile} symmetry indicated topological bands as those whose corresponding vectors $\mathbf{v}$ can only be expressed as a linear combination of EBR vectors with at least one negative coefficient.

Using this machinery, characterizing the vector space of symmetry-indicated topological band structures for a space group is a question regarding the existence of solutions to a Diophantine equation. Given a band structure specified by a vector $\mathbf{v}$, 
we ask whether $\mathbf{v}$ can be written as a linear combination of EBR vectors with integer coefficients. Collecting the vectors $\mathbf{e}_a$ for the EBRs into a matrix $
\mathbf{A}$, the question is to find an integer vector of EBR multiplicities $\mathbf{n}$ such that
\begin{equation}
    \mathbf{v}=\mathbf{A}\mathbf{n}.
\end{equation}
The space of vectors $\mathbf{v}$ for which a solution to these equations exists is found via the \emph{Smith decomposition} of the matrix $\mathbf{A}$. Given an integer-valued $n\times m$ matrix $\mathbf{A}$, the Smith decomposition writes $\mathbf{A}$ as
\begin{equation}
    \mathbf{A}=\mathbf{U^{-1}DV^{-1}}
\end{equation}
where $\mathbf{U}$ and $\mathbf{V}$ are invertible over the integers. $\mathbf{U}$ is $n \times n$ and $\mathbf{V}$ is $m\times m$. $\mathbf{D}$ is an $n\times m$ matrix known as the Smith normal form of $\mathbf{A}$. It is a non-negative integer matrix with zeros off the main diagonal, $D_{ij}=d_i\delta_{ij}$. The nonzero diagonal entries are known as the elementary divisors of $\mathbf{A}$, and can be thought of as an integer-valued analogy to singular values. 
Defining
\begin{align}
    \mathbf{v}'&=\mathbf{U}\mathbf{v} \\
    \mathbf{n}'&=\mathbf{V}^{-1}\mathbf{n}
\end{align}
our Diophantine system of equations reduces to finding solutions to
\begin{equation}
    \mathbf{v}'=\mathbf{D}\mathbf{n}'.
\end{equation}
{Integer solutions for $\mathbf{v}'$ only exist when $d_i$ divides $v'_i$. It follows that}
the space $\mathcal{V}$ of symmetry indicated topological bands is isomorphic to
\begin{equation}
    \mathcal{V}_G\approx \oplus_{i} \mathbb{Z}_{d_i},\label{eq:indicatorgroup}
\end{equation}
where $\mathbb{Z}_{d_i}$ is the group of integers modulo $d_i$ \cite{Po2017,song2017,elcoro2020application,po2020symmetry}.\footnote{This follows from the so-called structure theorem for finitely generated modules: the space of quasi-band reps can be written as a free module over the integers, even after incorporating the compatibility relations. The EBR matrix is a set of relations on this module, and Eq.~(\ref{eq:indicatorgroup}) follows from the structure theorem for finitely generated modules \cite{jacobson2012basic}.}
That this space is finite (i.e., contains no factors of $\mathbb{Z}$) was first demonstrated in Ref.~\cite{Po2017}, and is a mathematical reflection of the fact that symmetry eigenvalues can only determine integer-value invariants such as Chern numbers modulo the order of the symmetry \cite{Fang2012}. 
The module $\mathcal{V}_G$ is the ``symmetry indicator group'' for the space group $G$.

\subsubsection{Formulas for Symmetry Indicators}

Let us see how this works in practice using our example of the wallpaper group $p2$. Using Eqs.~(\ref{eq:ebrvec1}--\ref{eq:ebrvec2}), we form the {Smith decomposition of the EBR matrix. While $U$ and $V$ are gauge-dependent, we find for the elementary divisors}
\begin{equation}
    D=\left(\begin{array}{cccccccc}
    1 & 0 & 0 & 0 & 0 & 0 & 0 & 0 \\
    0 & 1 & 0 & 0 & 0 & 0 & 0 & 0 \\
    0 & 0 & 1 & 0 & 0 & 0 & 0 & 0 \\
    0 & 0 & 0 & 1 & 0 & 0 & 0 & 0 \\
    0 & 0 & 0 & 0 & 2 & 0 & 0 & 0 \\
    0 & 0 & 0 & 0 & 0 & 0 & 0 & 0 \\
    0 & 0 & 0 & 0 & 0 & 0 & 0 & 0 \\
    0 & 0 & 0 & 0 & 0 & 0 & 0 & 0 \\
    0 & 0 & 0 & 0 & 0 & 0 & 0 & 0 \\
    \end{array}\right)
\end{equation}
We see first that the distinct elementary divisors are $0,1,$ and $2$ with multiplicity $3,4,$ and $1$ respectively. The three zero divisors tell us that the eight EBRs provide an overcomplete generating set for the set of topologically trivial band structures. This we could see already from our expression Eq.~(\ref{eq:qbrp2}) for the general quasi-band representation, which depended on only five free parameters; the three elementary zero divisors correspond to the $8-5=3$ redundant band representations. The four elementary divisors equal to $1$ tell us that in the five-dimensional space of band structures, a four dimensional subspace can be generated entirely from EBRs with integer coefficients. The one nontrivial elementary divisor $d_i=2$ corresponds to the remaining subspace of band structures containing a topologically nontrivial element: 
this correpsonds to a $\mathbb{Z}_2$ symmetry indicator.

Using explicit forms for the matrices $U$ and $V$, we derive an expression for the symmetry indicator in terms of the representations appearing in a band structure \cite{song2017}. 
In this case, the indicator is given by $\nu = n_{\Gamma+}+n_{M+}-n_{X+}-n_{Y+} \mod 2 \in \mathbb{Z}_2$; it is impossible to choose an integer band representation vector $\mathbf{n}$ that gives a band structure with $\nu =  1$. We cast this in a familiar form by using the constraints of Eq.~(\ref{eq:qbrp2}) along with arithmetic modulo $2$ 
to find
\begin{equation}
    (-1)^\nu = \prod_{\mathrm{TRIM}} \zeta_i,
\end{equation}
i.e. that $(-1)^\nu$ is the product of all occupied parity eigenvalues. This is the famous Fu-Kane formula for the Chern number of a two-dimensional insulator \cite{Fu2007}. This formula was originally derived by considering the contraints that parity symmetry places on the distribution of Berry curvature throughout the BZ. Here, we see how the formula arises from the algebraic structure of band representations. Furthermore, using the systematic technique outlined here, similar formulas have been derived for all symmetry indicators in all 230 non-magnetic space groups, both with and without spin orbit coupling \cite{Po2017,song2017,elcoro2020application}. 
Symmetry indicators derived from topological quantum chemistry have led to the discovery of previously overlooked topological crystalline phases. Perhaps the most surprising of these has been the discovery of higher-order TIs.

\subsubsection{HOTIs}

Similar to our analysis of the wallpaper group $p2$, we can use TQC to derive the symmetry indicator formulas for the group $P\bar{1}1'$, {generated by three linearly independent lattice translations, inversion and time-reversal symmetry.} 
The full details of this derivation are presented in Refs.~\cite{khalaf,Po2017,elcoro2020application,song2017}; 
here, we present a quick proof of symmetry indicators beyond the strong $\mathbb{Z}_2$ invariant.
We first enumerate the elementary band representations in this space group. There are eight maximal Wyckoff positions $1a$--$1h$, corresponding to the eight inversion centers in the unit cell. The sixteen EBRs in this space group are obtained by induction from a Kramers pair of orbitals with either $+1$ or $-1$ inversion eigenvalue at one of these eight Wyckoff positions. The induction formulas in Sec.~\ref{sec:bandinduction} determine the little group representations at each of the eight time-reversal invariant momenta (TRIM) in the BZ: the eight Wyckoff positions have coordinates $1/2(n,m,\ell)$ where $n,m,\ell=0,1$. Furthermore, the translation vector $\mathbf{t}_{\alpha\beta}$ entering Eq.~(\ref{eq:rhoG3}) for the inversion representation matrix is twice the Wyckoff coordinate, $\mathbf{t}_{\alpha\beta}=(n,m,\ell)$. Thus, the inversion eigenvalues of states at momentum $\mathbf{k}_i$ in a band representation will match those of the position-space orbital if $\mathbf{k}_i\cdot\mathbf{t_{\alpha\beta}}=0$, and will be mismatched otherwise. For each EBR,  $\mathbf{k}_i\cdot\mathbf{t}_{\alpha\beta}=0$ at zero, four, or eight TRIM points. From this, we deduce that the number of TRIM points with negative inversion eigenvalue is $0 \mod 4$ for a band representation. 
{This observation suggests that the indicator group of $P\bar{1}1'$ is $\mathbb{Z}_4$, which is confirmed by the Smith decomposition.}
The corresponding $\mathbb{Z}_4$ index is given by \cite{khalaf}:
\begin{equation}
\nu_4= \sum_{\mathrm{TRIMS}\,\mathbf{k}_i}n_{\mathbf{k_i}-}\mod 4,
\end{equation}
where $n_{\mathbf{k_i}-}$ is the multiplicity of irreps (Kramer's pairs) at the TRIM $\mathbf{k}_i$ with negative inversion eigenvalue. 
The cases where $\nu_4=1,3$ are conventional strong TIs. 
The case where $\nu_4=2$ is topologically nontrivial, although the strong $\mathbb{Z}_2$ index vanishes. Furthermore, it is interesting to note that weak TIs have $\nu_4=2$, but while the weak indices require translational symmetry to be preserved, $\nu_4$ does not. Finally, there exist translationally invariant systems with $\nu_4=2$ and no nontrivial weak indices; these can be obtained by ``double band inversions" at a single TRIM point, by unit-cell doubling perturbations of a weak TI, or by increasing spin-orbit coupling in a monopole nodal-line semimetal \cite{fang2015topological,ahn2018band}. Such systems form the prototypical
example of (helical) \emph{higher order topological insulators} (HOTIs).

A distinguishing feature of HOTIs is the failure of the conventional bulk-boundary correspondence. Unlike the surfaces of strong TIs (or weak TIs with translation symmetry) two-dimensional surfaces of three-dimensional HOTIs do not have topologically protected surface states. However, on large rods or particles that respect the crystal symmetries, there are topologically protected one-dimensional states. 
For example, the HOTI protected by inversion and time-reversal in $P\bar{1}1'$ must have a single one-dimensional helical mode propagating along an inversion-symmetric arc on the surface of a finite system. 

While we have focused primarily on the case of space group $P\bar{1}1'$ for simplicity, symmetry-indicated higher-order topology is a ubiquitous phenomenon in crystals. Symmetry-indicated HOTIs can be found in any space group with inversion or rotoinversion symmetry \cite{khalaf}. By examining symmetry indicators, several candidate higher-order topological materials have been identified \cite{schindler2018higher,hotis,fang2020higher}, and suggestive experiments imaging hinge states in Bismuth crystals have been reinterpreted in light of new theoretical understanding \cite{schindler2018higher,hsu2019topology,nayak2019resolving}. Furthermore, by relaxing the constraints of time-reversal symmetry, chiral HOTIs with inversion symmetry have been discovered, such as the axion insulators MnBi$_2$Te$_4$ \cite{otrokov2019prediction,liu2020robust} and EuIn$_2$As$_2$ \cite{xu2019higher}.

Using the methods outlined here, several complete classifications of symmetry-indicated phases and materials have been compiled. However, the applicability of TQC extends beyond symmetry indicators. Next, we will see how the position-space approach to band representations and topology allows us to diagnose topological crystalline phenomena absent symmetry indicators.

\section{Beyond Symmetry Indicators}\label{sec:beyond}

A nontrivial value for a symmetry indicator is a sufficient condition for a group of bands to be topologically nontrivial, but it is not a necessary condition. In many cases, bands may be topologically nontrivial despite having little group irreps that are identical to a sum (or difference) of EBRs. This occurs, for instance, for strong TIs and Chern insulators in space groups without inversion or rotoinversion symmetries. Recently, it has even been shown that there exist classes of HOTIs with no symmetry indicators \cite{fangrotational,fang2019new}. 
We will now show how TQC can shed light on topological bands beyond momentum-space irreps. We will see how the lack of exponentially-localized Wannier functions for topological bands emerges from a position-space picture. To start, we first show how certain space groups admit trivial atomic insulators that are distinguishable in position space, but indistinguishable in momentum space.

\subsection{Disconnected EBRs}\label{sec:disconnected}

Recall that EBRs form the fundamental building blocks of trivial band structures: all topologically trivial electronic bands in crystals are equivalent to a sum of EBRs. In position space, this sum reflects the symmetry properties of the exponentially localized Wannier functions describing the electronic states; in momentum space, the sum gives the little group representations under which the wavefunctions transform at every $\mathbf{k}$. However, a priori there is no connection between the little group representations in an EBR and the connectivity of the electronic bands in an EBR. While the compatibility relations of Sec.~\ref{sec:compatibility} constrain the connectivity of EBR bands, they do not require that energy bands in an EBR must always be connected. 

Let us suppose that the compatibility relations allow for the bands transforming in an EBR $\rho_\mathbf{k}$ to be disconnected in the BZ. In this case, the EBR $\rho_\mathbf{k}$ can be written as a sum of quasi-band representations $\rho_\mathbf{k}=\rho_{1\mathbf{k}}\oplus\rho_{2\mathbf{k}}$. If $\rho_{1\mathbf{k}}$ and $\rho_{2\mathbf{k}}$ are both band representations, then $\rho_\mathbf{k}$ is an EBR that is the sum of two band representations, which contradicts the definition of an EBR from Sec.~\ref{sec:EBR}. Thus, we conclude that at least one of the quasi-band representations $\rho_{1\mathbf{k}}$ and $\rho_{2\mathbf{k}}$ cannot be a band representation, and so one of the two must be topologically nontrivial. There are then three possibilities: First, it may be that $\rho_{1\mathbf{k}}$ is a band representation, while $\rho_{2\mathbf{k}}$ is not; we will analyze this situation in detail in Sec.~\ref{sec:fragile}. Second, it is possible that although the compatibility relations are solved, there are generic band crossings at other points in the BZ, leading to a topological semimetallic phase. Here, we will focus on the third possibility, where neither $\rho_{1\mathbf{k}}$ nor $\rho_{1\mathbf{k}}$ are band representations, and hence both sets of disconnected bands are topologically nontrivial. In this case, the bands transforming in the quasi-band representations $\rho_{1\mathbf{k}}$ and $\rho_{2\mathbf{k}}$ cannot be described by exponentially localized Wannier functions. 

Note that our argument did not make any reference to the little group representations in the quasi-band representations $\rho_{1\mathbf{k}}$ and $\rho_{2\mathbf{k}}$. Provided that the compatibility relations are satisfied, we were able to deduce the existence of topologically nontrivial bands (or a topological semimetal) using only TQC. In particular, it is possible that the quasi-band representations $\rho_{1\mathbf{k}}$ and $\rho_{2\mathbf{k}}$ share the same little group representations with a (sum of) EBRs. Nevertheless, TQC encodes the nontrivial nature of these bands in the momentum-space structure of the quasi-band representation. 

As an example, we analyze the Kane-Mele model of spin-orbit coupled graphene \cite{Kane04,kane2005z} from the perspective of TQC, following Ref.~\cite{NaturePaper}. The Kane-Mele model consists of a four band EBR, induced from spinful $p_z$ orbitals at the honeycomb sites ($2b$ Wyckoff position) in wallpaper group $p6mm$. In the topological phase, this EBR splits into two disconnected components, each of which has a nontrivial $\mathbb{Z}_2$ index.
Because these two groups of bands come from a disconnected elementary band representation, at least one of the valence or conduction band must be topologically nontrivial; in this example, both are. 
Interestingly, only one component yields little group irreps inconsistent with a sum of EBRs. 
Thus, for the other group of bands, their nontrivial topology is entirely hidden from detection by little group irreps.

\subsection{Hidden Obstructed Atomic Limits}

Let us consider two groups of bands with the same irreps at all high-symmetry points in the BZ, which we refer to as \definition{irrep-equivalent}.
It was pointed out in Refs.~\cite{Bacry1988b,Michel1992} that it is possible that two such bands {that transform identically under all symmetries at each point in the BZ differ by a topologically non-trivial} global gauge transformation, thus rendering them distinct.
These bands do not need to be topological: in more than one spatial dimension, distinct trivial phases can be irrep-equivalent, but distinguished by topological invariants.
(In one dimension, the only crystal symmetry is inversion, which completely distinguishes distinct phases.)

As an example of irrep-equivalent EBRs that are not equivalent, in space group $F222$, for each EBR induced from the $4a$ position, there is an irrep-equivalent EBR induced from the $4b$ position \cite{Bacry1988b,Michel1992,Bacry1993,EBRTheoryPaper}. In the language of TQC, we explain this result by examining the matrices $\rho_\mathbf{k}$ for the different band representations in this space group. In fact, using a basis function $|w_a\rangle$ at the $4a$ position in $F222$, a basis for a representation of the site-symmetry group of the $4b$ position is given via the gauge transformation
\begin{equation}
|w_b\rangle = \exp(i(k_2+k_3-k_1)/2)|w_a\rangle.
\label{eq:F222unitary}
\end{equation}
Since irrep characters are invariant under unitary transformations, Eq.~(\ref{eq:F222unitary}) implies that every band representation at the $4b$ position must be irrep equivalent to a band representation at the $4a$ position. However, the unitary transformation in Eq.~(\ref{eq:F222unitary}) is not periodic in the BZ. It thus cannot arise from any smooth and periodic deformation of a band representation at the $4a$ position, and hence does not represent an equivalence as per Sec.~\ref{sec:EBR}. 
Thus, even though band representations at $4a$ and $4b$ are irrep equivalent, they are not equivalent.
Consequently, a parametric family of Hamiltonians $H(t)$ with occupied bands realizing the $4a$ band representation at $t=0$ and the $4b$ band representation at $t=1$ must pass through a gap-closing phase transition for some intermediate $t$;
hence we refer to the two band representations as \definition{obstructed atomic limits} (OALs) \cite{NaturePaper,EBRTheoryPaper}, which, in this case, are hidden from a symmetry-indicated diagnosis.
This hidden obstructed atomic limit does not have a quantized polarization \cite{Michel1992}; there is, however, a product of Berry phases (and curvatures \cite{hiddenpaper,zeiner1996nonlinear}) that can be used to distinguish these two atomic limits. 
Thus there exist distinct trivial phases that can only be distinguished through the analytic properties of band representations. In the next subsection, we will see how hidden OALs can be used as the building blocks of a new family of TCIs beyond symmetry indicators.

\subsection{Rotational anomaly insulators}
\label{sec:rotationalanomaly}

In three dimensions there is another large class of TIs beyond the symmetry indicator paradigm, protected by rotational and time reversal symmetries \cite{fang2019new}. In two-dimensional systems with twofold rotation and time-reversal symmetry, the Berry phase of electronic states along any closed, symmetric path is real \cite{Fang2012}. Rotational symmetry then requires that a two-dimensional system with $n$-fold rotational symmetry must have $0\mod 2n$ two-band crossings (Dirac cones) at the Fermi level when $n$ is even. A Hamiltonian with even $n$-fold rotation and time reversal symmetry and $n$ Dirac cones cannot be realized in an isolated two-dimensional system.

Consider then a three dimensional insulator with an $n$-fold rotation axis, {with $n$ even}, and a surface that preserves the rotational symmetry. This surface may host $n$ Dirac cones at the Fermi level. By the preceding discussion, these surface states cannot be removed by coupling to any additional two-dimensional degrees of freedom. In fact, these surface states emerge from a topologically nontrivial bulk phase protected by $n$-fold rotation symmetry, which is generically not symmetry indicated. The bulk topology can be understood in several ways. In momentum space, we can consider a three dimensional bulk Hamiltonian $H(k_1,k_2,k_3)$ as a family of rotationally-invariant two-dimensional Hamiltonians $H_{k_3}(k_1,k_2)$ parametrized by $k_3$, the momentum along the rotation axis. Absent any weak topological order, At $k_3=0$ and $k_3=\pi$ this Hamiltonian gives a two-dimensional insulator in a time-reversal and rotationally symmetric atomic limit, with Wannier functions pinned to a maximal Wyckoff position of the rotation group. In wallpaper groups $p2$, $p4$, and $p6$ there are composite band representations from different maximal Wyckoff positions that are indistinguishable from their little group irreps, i.e., hidden OALs. The rotational anomaly insulators are three dimensional Hamiltonians that interpolate between different hidden atomic limits \cite{hiddenpaper}. 

We can also construct the rotational anomaly insulators from a position-space perspective. 
Recall that periodically stacked two-dimensional quantum spin-Hall insulators with appropriate coupling yield a surface theory with an even number of Dirac cones \cite{fukanemele}. Similarly, we obtain a rotational anomaly insulator by layering quantum spin-Hall insulators while simultaneously preserving translation symmetry and rotational symmetry. The $n$-fold rotationally symmetric insulator can be shown to be equivalent to $n$ stacks of quantum spin-Hall insulators, where the stacking vectors transform into each other under the rotational symmetry; this gives a periodic array of intersecting quantum-spin Hall insulators, each of which contains a rotation axis. At each rotation axis, $n/2$ quantum spin-Hall insulators intersect. 

While this construction uses translation symmetry in an essential way, the rotational anomaly insulator remains nontrivial if translation symmetry is broken, provided rotational symmetry is preserved. With only rotational symmetry, the $n$ surface Dirac cones at the boundary of the system can gap, leaving behind $n$ hinge states. Thus, without translational symmetry the rotational anomaly insulators become non-symmetry indicated HOTIs.

\section{Fragile Topology}\label{sec:fragile}

In Sec.~\ref{sec:disconnected}, we showed that if an EBR consists of disconnected sets of energy bands in the BZ, then at least one of those sets of bands must be topologically nontrivial. 
We considered in detail the case where both bands are nontrivial, both with and without symmetry indicators. However, we tacitly assumed that the topology was stable to the addition of trivial bands. As was first noted by Ref.~\cite{Fragile2017}, stability is violated when only one of two sets of disconnected bands is topologically nontrivial. In these situations, the topological bands are said to exhibit ``fragile topology.'' The distinguishing feature of a fragile topological band is that it may be combined with a trivial band to produce a trivial set of bands. For disconnected EBRs with one trivial and one topological band, this is immediately clear: the topological band, combined with the trivial band, recreate the original EBR, which is trivial.

To find an example of fragile topology, we revisit the Kane-Mele model. By adding sufficiently long-range spin-dependent hopping, the split elementary band representation discussed in Sec.~\ref{sec:disconnected} need not possess a nontrivial $\mathbb{Z}_2$ invariant \cite{Fragile2017,Slager2018,bradlyn2019disconnected}. In fact, for judicious choices of parameters, 
{the valence bands support exponentially localized Wannier functions centered at the $1a$ Wyckoff position, while the conduction bands are (fragile) topological; the little group irreps cannot be obtained from a sum of EBRs and they have nontrivial Berry phases \cite{bradlyn2019disconnected,Slager2018,Fragile2017}.}

As with stable topological bands, fragile topology may be symmetry indicated or ``hidden." In the language of Sec.~\ref{sec:symmetryindicators}, symmetry indicated fragile topology occurs when the vector of irrep multiplicities $\mathbf{v}$ for a given set of bands can be expressed as a linear combination of EBR vectors requiring at least one negative coefficient, 
\begin{equation}
    \mathbf{v}=\sum_{a+} n_{a+}\mathbf{e}_{a+} - \sum_{a-} n_{a-}\mathbf{e}_{a-}\label{eq:fragiledecomp}
\end{equation}
When this occurs, all stable symmetry indicators are by definition trivial (since a negative integer is still an integer). However, because any trivial band structure must consist of a positive sum of EBRs, the vector $\mathbf{v}$ cannot correspond to a trivial band structure. 
Thus, if $\mathbf{v}$ does not have a non-symmetry indicated stable topology, it corresponds to a fragile topological band structure, which can be trivialized by adding to it all EBRs $\mathbf{e}_{a-}$ that have negative coefficients in the decomposition Eq.~(\ref{eq:fragiledecomp}). As with the classification of symmetry indicators for stable topology, the set of symmetry indicated fragile topological bands in all space groups have been tabulated \cite{song2020twisted,song2019fragile}. The classification scheme requires finding solutions to the Diophantine equations $\mathbf{v}=\mathbf{A}\mathbf{n}$ which cannot be expressed with all positive integer coefficients. 

The set of symmetry indicators for fragile topology is orders of magnitude larger than for stable topology. This is one indication of the ubiquity of fragile topology in nature. Fragile topology has been shown to play a crucial role in the low-energy physics of twisted bilayer graphene, where it is responsible for the anomalous localization of the flat-band Wannier functions relevant for strongly correlated physics  \cite{po2018origin,zou2018band,song2018all,ahn2019failure}. 
Similarly, fragile topology is prevalent in systems with combined twofold rotation and time-reversal symmetry in two dimensions.

\subsection{Real structures and $C_2T$}
\label{sec:real}

In two-dimensions, fragile topology often manifests in systems with $C_2T$ symmetry, where $C_2T$ is the composition of a twofold rotation and time-reversal symmetry. In momentum space, $C_2T$ leaves every $\mathbf{k}$-point invariant, and so can be represented as a local antiunitary operation on Bloch states. 
Examining the square of this symmetry:
\begin{equation}
(C_2T)^2=C_2TC_2T=C_2^2T^2=\bar{E}^2=E,
\end{equation}
we find that in both single (spinless) and double (spin-1/2) valued representations, $C_2T$ squares to the identity. In a system with $C_2T$ symmetry and $N$ total bands, an $N\times N$ matrix representation of $C_2T$ is furnished by $\Delta(C_2T)=J(\mathbf{k})\mathcal{K}$, where $\mathcal{K}$ is the operation of complex conjugation, and the matrix $J(\mathbf{k})$ is a smooth function of $\mathbf{k}$. This defines a ``real structure'' on the basis of Bloch states at every point in $\mathbf{k}$, and we can write the Hilbert space $\mathcal{H}$ in terms of a \emph{real} Hilbert space 
\begin{equation}
\mathcal{H}_\mathbb{R}\equiv\left\{ \frac{1}{2}(|a_{n\mathbf{k}}\rangle+\Delta(C_2T) | a_{n\mathbf{k}}\rangle) \Big| |a_{n\mathbf{k}}\rangle\in\mathcal{H}\right\}
\end{equation}
as
\begin{equation}
\mathcal{H}=\mathbb{C}\otimes\mathcal{H}_\mathbb{R}
\end{equation}
In words, $C_2T$ identifies a canonical basis of real-valued functions for the Hilbert space $\mathcal{H}$ with the basis for the Hilbert space $\mathcal{H}_\mathbb{R}$. Since the Hamiltonian $H$ is $C_2T$ symmetric, in this basis the Hamiltonian must be real at every $\mathbf{k}$. Projection operators onto occupied bands inherit this real structure as well; the occupied band Bloch functions $\{u_{n\mathbf{k}}\}$, $n=1,\dots,N_{\mathrm{occ}}$ span a real vector space at each $\mathbf{k}$, and so form a ``real vector bundle'' over the BZ \cite{fang2015topological,ahn2018band}.

The importance of this formalism stems from a well-known fragile topological invariant characterizing real vector bundles in two dimensions: the Euler class. The Euler class is the generalization of the Euler characteristic of a manifold to vector bundles other than the tangent bundle \cite{nakahara2003geometry}. It can be regarded as the winding number $\nu$ of the orthogonal transformations necessary to define the real basis for the Bloch states $\{u_{n\mathbf{k}}\}$ at all $\mathbf{k}$ in the BZ (provided the electrical polarization vanishes) \cite{ahn2018higher,ahn2019failure,ahn2019stiefel}. If $N_{\mathrm{occ}}=2$, then this winding can be expressed concretely in terms of the Berry curvature. With $C_2T$ symmetry, the Berry connection preserves the real structure of the vector space, and hence is an $SO(N_{\mathrm{occ}})=SO(2)$ matrix. The Berry curvature $\Omega_{nm}$, then, is a single number
\begin{equation}
\Omega_{nm}=\Omega\epsilon_{nm},
\end{equation}
where $\epsilon_{nm}$ is the Levi-Civita symbol. The fragile Euler invariant is given, in analogy with the Gauss-Bonnet theorem, as
\begin{equation}
\nu=\frac{1}{2\pi}\int d^2k \Omega(\mathbf{k}) \in \mathbb{Z}.
\end{equation}
(In fact, this is an example of the more general Chern-Gauss-Bonnet theorem for real vector bundles \cite{bell2006gauss}). A nonzero value for the two-band Euler class $\nu$ indicates the inability to define a real basis for the occupied bands globally over the entire BZ. This manifests as a winding in the non-abelian Berry phase for the two occupied bands \cite{bradlyn2019disconnected,Slager2018}. This winding is similar to a quantum spin-Hall insulator, where the winding of the Berry phase indicates an obstruction to finding a globally defined time-reversal symmetric basis for the occupied bands. However, unlike the quantum spin Hall invariant, the Euler class can be trivialized by adding an additional trivial occupied band; it is a fragile invariant. Since $C_2T$ is an antiunitary symmetry, band structures with different Euler classes cannot in general be distinguished by little group representations at high-symmetry points.


Although the Euler class is fragile, the parity of the Euler class
\begin{equation}
w_2=\nu\mod 2
\end{equation}
is a stable invariant, known as the second Stiefel Whitney class of the vector bundle of Bloch states. In general $w_2=1$ is indicative of an OAL, similar to the electrical polarization (which is also quantized modulo 2 with $C_2T$, and coincides with the first Stiefel Whitney number $w_1$).

\section{Applications and Outlook}\label{sec:outlook}


One {breakthrough application} of topological quantum chemistry is the rapid screening for new topological materials. 
A typical strategy amenable to high-throughput computations is to use ab-initio methods to compute the little group representations for the occupied bands in a material, check that they satisfy the compatibility relations and compute symmetry indicators. A failure to satisfy the compatibility relations indicates the presence of a topological semimetal: the occupied bands must connect to the unoccupied bands along some path in the BZ. On the other hand, a nontrivial symmetry indicator reflects the presence of a topological insulator or semimetal. This approach has been leveraged at large scales to filter materials databases for candidate topological materials \cite{bigmaterials,bigmaterials-china,ashvin-materials,tqcsite,he2019symtopo}.

Beyond high-throughput screening, TQC is {a powerful tool} to target space groups and crystal structures prone to topological behavior by searching for disconnected EBRs. This is particularly useful in the design of topological metamaterials \cite{peri2020experimental,de2019engineering,alexandradinata2019crystallographic}, where one has precise control over the EBRs appearing in the spectrum for a given structure. This has enabled, for instance, tunable experimental control over fragile topological bands.

Looking towards the future, there are several avenues for further development and application of TQC. First, TQC has so far been applied primarily to time-reversal invariant crystals. One path forward is the extension to magnetic symmetry groups, enabling a full catalogue of band topology in commensurate magnetic systems \cite{watanabe2018structure,xu2020high}. Additionally, the position space perspective of TQC can be brought to bear on the role of disorder in TCIs \cite{song2020twisted}. Finally, the theory of TCIs outlined here is applicable only insofar as a single-electron picture is a good description. The tools of TQC, however, can shed light on how electron-electron correlations alter the behavior of topological materials; several materials identified as topologically interesting in the single-electron regime have already proven to be even more intriguing when correlations are taken into account \cite{gooth2019evidence,shi2019charge,dmft,rachel2018interacting,po2018origin,zou2018band,song2018all,ahn2019failure,dzero2016topological,khalaf2018higher,langbehn2017reflection,wang2018weak,bultinck2019three,ono2019symmetry,schindler2020pairing}. 

\section*{ACKNOWLEDGMENTS}

We acknowledge many informative conversations with Mois Aroyo, Andrei Bernevig, Luis Elcoro, Maia Garcia Vergniory and Benjamin Wieder during previous collaborations.
J.C. is partially supported by the Flatiron Institute, a division of the Simons Foundation.

\bibliographystyle{ar-style4}
\bibliography{refs}

\begin{thebibliography}{125}
\expandafter\ifx\csname natexlab\endcsname\relax\def\natexlab#1{#1}\fi

\bibitem{klitzing1980new}
Klitzing Kv, Dorda G, Pepper M. 1980.
\textit{Physical Review Letters} 45:494

\bibitem{laughlin1981quantized}
Laughlin RB. 1981.
\textit{Physical Review B} 23:5632

\bibitem{tsui1982two}
Tsui DC, Stormer HL, Gossard AC. 1982.
\textit{Physical Review Letters} 48:1559

\bibitem{laughlin1983anomalous}
Laughlin RB. 1983.
\textit{Physical Review Letters} 50:1395

\bibitem{Kane04}
Kane CL, Mele EJ. 2005{\natexlab{a}}.
\textit{Phys. Rev. Lett.} 95:226801

\bibitem{bernevig2006quantum}
Bernevig BA, Hughes TL, Zhang SC. 2006.
\textit{Science} 314:1757--1761

\bibitem{konig2007quantum}
K{\"o}nig M, Wiedmann S, Br{\"u}ne C, Roth A, Buhmann H, et~al. 2007.
\textit{Science} 318:766--770

\bibitem{fukanemele}
Fu L, Kane CL, Mele EJ. 2007.
\textit{Phys. Rev. Lett.} 98:106803

\bibitem{xia2009observation}
Xia Y, Qian D, Hsieh D, Wray L, Pal A, et~al. 2009.
\textit{Nature physics} 5:398

\bibitem{Fu2011}
Fu L. 2011.
\textit{Phys. Rev. Lett.} 106:106802

\bibitem{Hsieh2012}
Hsieh TH, Lin H, Liu J, Duan W, Bansil A, Fu L. 2012.
\textit{Nature Communications} 3:932

\bibitem{Turner2010}
Turner AM, Zhang Y, Vishwanath A. 2010.
\textit{Phys. Rev. B} 82:241102

\bibitem{Turner2012}
Turner AM, Zhang Y, Mong RSK, Vishwanath A. 2012.
\textit{Phys. Rev. B} 85:165120

\bibitem{Hourglass}
Wang Z, Alexandradinata A, Cava RJ, Bernevig BA. 2016.
\textit{Nature} 532:189--194

\bibitem{Wieder17}
Wieder BJ, Bradlyn B, Wang Z, Cano J, Kim Y, et~al. 2017

\bibitem{hotis}
Schindler F, Cook AM, Vergniory MG, Wang Z, Parkin SSP, et~al.
  2018{\natexlab{a}}.
\textit{Science Advances} 4

\bibitem{benalcazar2017quantized}
Benalcazar WA, Bernevig BA, Hughes TL. 2017.
\textit{Science} 357:61--66

\bibitem{fangrotational}
Song Z, Fang Z, Fang C. 2017.
\textit{Phys. Rev. Lett.} 119:246402

\bibitem{Aris2014}
Alexandradinata A, Dai X, Bernevig BA. 2014.
\textit{Phys. Rev. B} 89:155114

\bibitem{Fang2012}
Fang C, Gilbert MJ, Bernevig BA. 2012.
\textit{Phys. Rev. B} 86:115112

\bibitem{Hughes11}
Hughes TL, Prodan E, Bernevig BA. 2011.
\textit{Phys. Rev. B} 83:245132

\bibitem{xu2015discovery}
Xu SY, Belopolski I, Alidoust N, Neupane M, Bian G, et~al. 2015.
\textit{Science} 349:613--617

\bibitem{Lv15}
Lv BQ, Xu N, Weng HM, Ma JZ, Richard P, et~al. 2015{\natexlab{a}}.
\textit{Nat. Phys.}

\bibitem{Lv15a}
Lv BQ, Weng HM, Fu BB, Wang XP, Miao H, et~al. 2015{\natexlab{b}}.
\textit{Phys. Rev. X} 5:031013

\bibitem{Young2015}
Young SM, Kane CL. 2015.
\textit{Phys. Rev. Lett.} 115:126803

\bibitem{Steinberg14}
Steinberg JA, Young SM, Zaheer S, Kane CL, Mele EJ, Rappe AM. 2014.
\textit{Phys. Rev. Lett.} 112:036403

\bibitem{Young12}
Young SM, Zaheer S, Teo JCY, Kane CL, Mele EJ, Rappe AM. 2012.
\textit{Phys. Rev. Lett.} 108:140405

\bibitem{Bradlyn2016}
Bradlyn B, Cano J, Wang Z, Vergniory MG, Felser C, et~al. 2016.
\textit{Science} :10.1126/science.aaf5037

\bibitem{schroter2019chiral}
Schr{\"o}ter NBM, Pei D, Vergniory MG, Sun Y, Manna K, et~al. 2019.
\textit{Nature Physics} :1

\bibitem{Rao2019}
Rao Z, Li H, Zhang T, Tian S, Li C, et~al. 2019.
\textit{Nature} 567:496--499

\bibitem{Sanchez2018}
Sanchez DS, Belopolski I, Cochran TA, Xu X, Yin JX, et~al. 2019.
\textit{Nature} 567:500--505

\bibitem{Wan11}
Wan X, Turner AM, Vishwanath A, Savrasov SY. 2011.
\textit{Phys. Rev. B} 83:205101

\bibitem{TangEA17}
Tang P, Zhou Q, Zhang SC. 2017.
\textit{Phys. Rev. Lett.} 119:206402

\bibitem{Shiozaki2014}
Shiozaki K, Sato M. 2014.
\textit{Phys. Rev. B} 90:165114

\bibitem{Chiu2013}
Chiu CK, Yao H, Ryu S. 2013.
\textit{Phys. Rev. B} 88:075142

\bibitem{chiu2014}
Chiu CK, Schnyder AP. 2014.
\textit{Phys. Rev. B} 90:205136

\bibitem{ando2015topological}
Ando Y, Fu L. 2015.
\textit{Annu. Rev. Condens. Matter Phys.} 6:361--381

\bibitem{NaturePaper}
Bradlyn B, Elcoro L, Cano J, Vergniory MG, Wang Z, et~al. 2017.
\textit{Nature} 547:298--305

\bibitem{EBRTheoryPaper}
Cano J, Bradlyn B, Wang Z, Elcoro L, Vergniory MG, et~al. 2018.
\textit{Phys. Rev. B} 97:035139

\bibitem{GraphTheoryPaper}
Bradlyn B, Elcoro L, Vergniory MG, Wang Z, Cano J, et~al. 2018.
\textit{Phys. Rev. B} 97:035137

\bibitem{Marzari2012}
Marzari N, Mostofi AA, Yates JR, Souza I, Vanderbilt D. 2012.
\textit{Rev. Mod. Phys.} 84:1419

\bibitem{Zak1980}
Zak J. 1980.
\textit{Phys. Rev. Lett.} 45:1025--1028

\bibitem{Zak1981}
Zak J. 1981.
\textit{Physical Review B} 23:2824--2835

\bibitem{Zak1982}
Zak J. 1982.
\textit{Phys. Rev. B} 26:3010

\bibitem{Michel2001}
Michel L, Zak J. 2001.
\textit{Phys. Rep.} 341:377

\bibitem{comment}
Po HC, Watanabe H, Vishwanath A. 2018.
\textit{Phys. Rev. Lett.} 121:126402

\bibitem{Slager2018}
Bouhon A, Black-Schaffer AM, Slager RJ. 2019.
\textit{Phys. Rev. B} 100:195135

\bibitem{bradlyn2019disconnected}
Bradlyn B, Wang Z, Cano J, Bernevig BA. 2019.
\textit{Physical Review B} 99:045140

\bibitem{Fragile2017}
{Cano} J, {Bradlyn} B, {Wang} Z, {Elcoro} L, {Vergniory} MG, et~al. 2018.
\textit{Phys. Rev. Lett.} 120:266401

\bibitem{bigmaterials}
Vergniory M, Elcoro L, Felser C, Bernevig B, Wang Z. 2019.
\textit{Nature} 566:480--485

\bibitem{bigmaterials-china}
Zhang T, Jiang Y, Song Z, Huang H, He Y, et~al. 2019.
\textit{Nature} 566:475--479

\bibitem{ashvin-materials}
Tang F, Po HC, Vishwanath A, Wan X. 2019.
\textit{Nature Physics} 15:470--476

\bibitem{Cracknell}
Bradley CJ, Cracknell AP. 1972.
The mathematical theory of symmetry in solids.
Oxford: Clarendon Press

\bibitem{ITA}
Aroyo MI, ed. 2016.
International tables for crystallography.
vol.~A.
International Union of Crystallography

\bibitem{Bilbao1}
Aroyo MI, Perez-Mato JM, Orobengoa D, Tasci E, de~la Flor G, Kirov A. 2011.
\textit{Bulg. Chem. Commun.} 43(2):183

\bibitem{Bilbao2}
Aroyo MI, Perez-Mato JM, Capillas C, Kroumova E, Ivantchev S, et~al.
  2006{\natexlab{a}}.
\textit{Z. Krist.} 221:15

\bibitem{Bilbao3}
Aroyo MI, Kirov A, Capillas C, Perez-Mato JM, Wondratschek H.
  2006{\natexlab{b}}.
\textit{Acta Cryst.} A62:115

\bibitem{Bilbao4}
Aroyo MI, Orobengoa D, de~la Flor G, Tasci ES, Perez-Mato JM, Wondratschek H.
  2014.
\textit{Acta Crystallographica Section A} 70:126--137

\bibitem{michel2001fundamental}
Michel L. 2001.
\textit{Physics Reports} 341:265--336

\bibitem{Kittel87}
Kittel C. 1987.
Quantum theory of solids.
Wiley

\bibitem{ashcroft2005solid}
Ashcroft NW, Mermin ND. 2005.
\textit{Google Scholar} 403

\bibitem{Zakphase}
Zak J. 1989.
\textit{Phys. Rev. Lett.} 62:2747

\bibitem{ksv}
King-Smith RD, Vanderbilt D. 1993.
\textit{Phys. Rev. B} 47:1651(R)

\bibitem{Serre}
Serre JP. 1996.
Linear representations of finite groups.
Springer

\bibitem{ArisCohomology}
Alexandradinata A, Wang Z, Bernevig BA. 2016.
\textit{Phys. Rev. X} 6:021008

\bibitem{Bacry1988b}
Bacry H, Michel L, Zak J. 1988.
\textit{Phys. Rev. Lett.} 61:1005--1008

\bibitem{Michel1992}
Michel L, Zak J. 1992.
\textit{EPL (Europhysics Letters)} 18:239

\bibitem{Bacry1993}
Bacry H. 1993.
\textit{Commun. Math. Phys.} 153:359

\bibitem{hiddenpaper}
Cano J, Bradlyn B, Elcoro L, Aroyo MI, Bernevig BA. 2020.
In prep

\bibitem{progbandrep}
\textrm{Bilbao Crystallogr. Server}. 2017.
Bandrep: Band representations of the double space groups.
\url{http://www.cryst.ehu.es/cgi-bin/cryst/programs/bandrep.pl}

\bibitem{GroupTheoryPaper}
Elcoro L, Bradlyn B, Wang Z, Vergniory MG, Cano J, et~al. 2017.
\textit{J. Appl. Cryst.} 50:1457--1477

\bibitem{ryu2010topological}
Ryu S, Schnyder AP, Furusaki A, Ludwig AW. 2010.
\textit{New Journal of Physics} 12:065010

\bibitem{schnyder2008classification}
Schnyder AP, Ryu S, Furusaki A, Ludwig AW. 2008.
\textit{Physical Review B} 78:195125

\bibitem{KitaevClassify}
{Kitaev} A. 2009.
{Periodic table for topological insulators and superconductors}. In
  \textit{American Institute of Physics Conference Series}, eds. V~{Lebedev},
  M~{Feigel'Man}, vol. 1134 of \textit{American Institute of Physics Conference
  Series}

\bibitem{balents2007topological}
Moore JE, Balents L. 2007.
\textit{Phys. Rev. B} 75:121306

\bibitem{roy2009topological}
Roy R. 2009.
\textit{Phys. Rev. B} 79:195322

\bibitem{wieder2018axion}
Wieder BJ, Bernevig BA. 2018.
\textit{arXiv preprint arXiv:1810.02373}

\bibitem{Po2017}
Po HC, Vishwanath A, Watanabe H. 2017.
\textit{Nat. Comm.} 8:50

\bibitem{witten2016fermion}
Witten E. 2016.
\textit{Reviews of Modern Physics} 88:035001

\bibitem{po2020symmetry}
Po HC. 2020.
\textit{Journal of Physics: Condensed Matter}

\bibitem{song2017}
Song Z, Zhang T, Fang Z, Fang C. 2018.
\textit{Nat. Commun.} 9:3530

\bibitem{elcoro2020application}
Elcoro L, Song Z, Bernevig BA. 2020.
\textit{arXiv preprint arXiv:2002.03836}

\bibitem{jacobson2012basic}
Jacobson N. 2012.
Basic algebra i.
Courier Corporation

\bibitem{Fu2007}
Fu L, Kane CL. 2007.
\textit{Phys. Rev. B} 76:045302

\bibitem{khalaf}
Khalaf E, Po HC, Vishwanath A, Watanabe H. 2018.
\textit{Phys. Rev. X} 8:031070

\bibitem{fang2015topological}
Fang C, Chen Y, Kee HY, Fu L. 2015.
\textit{Physical Review B} 92:081201

\bibitem{ahn2018band}
Ahn J, Kim D, Kim Y, Yang BJ. 2018.
\textit{Physical review letters} 121:106403

\bibitem{schindler2018higher}
Schindler F, Wang Z, Vergniory MG, Cook AM, Murani A, et~al.
  2018{\natexlab{b}}.
\textit{Nature physics} 14:918--924

\bibitem{fang2020higher}
Fang Y, Cano J. 2020.
\textit{Phys. Rev. B} 101:245110

\bibitem{hsu2019topology}
Hsu CH, Zhou X, Chang TR, Ma Q, Gedik N, et~al. 2019.
\textit{Proceedings of the National Academy of Sciences} 116:13255--13259

\bibitem{nayak2019resolving}
Nayak AK, Reiner J, Queiroz R, Fu H, Shekhar C, et~al. 2019.
\textit{Science advances} 5:eaax6996

\bibitem{otrokov2019prediction}
Otrokov M, Klimovskikh I, Bentmann H, Estyunin D, Zeugner A, et~al. 2019.
\textit{Nature} 576:416--422

\bibitem{liu2020robust}
Liu C, Wang Y, Li H, Wu Y, Li Y, et~al. 2020.
\textit{Nature Materials} :1--6

\bibitem{xu2019higher}
Xu Y, Song Z, Wang Z, Weng H, Dai X. 2019.
\textit{Physical review letters} 122:256402

\bibitem{fang2019new}
Fang C, Fu L. 2019.
\textit{Science Advances} 5:eaat2374

\bibitem{kane2005z}
Kane CL, Mele EJ. 2005{\natexlab{b}}.
\textit{Physical review letters} 95:146802

\bibitem{zeiner1996nonlinear}
Zeiner P, Dirl R, Davies B. 1996.
\textit{Physical Review B} 54:2466

\bibitem{song2020twisted}
Song ZD, Elcoro L, Bernevig BA. 2020.
\textit{Science} 367:794--797

\bibitem{song2019fragile}
Song Z, Elcoro L, Regnault N, Bernevig BA. 2019{\natexlab{a}}.
\textit{arXiv preprint arXiv:1905.03262}

\bibitem{po2018origin}
Po HC, Zou L, Vishwanath A, Senthil T. 2018.
\textit{Physical Review X} 8:031089

\bibitem{zou2018band}
Zou L, Po HC, Vishwanath A, Senthil T. 2018.
\textit{Physical Review B} 98:085435

\bibitem{song2018all}
Song Z, Wang Z, Shi W, Li G, Fang C, Bernevig BA. 2019{\natexlab{b}}.
\textit{Phys. Rev. Lett.} 123:036401

\bibitem{ahn2019failure}
Ahn J, Park S, Yang BJ. 2019.
\textit{Physical Review X} 9:021013

\bibitem{nakahara2003geometry}
Nakahara M. 2003.
Geometry, topology and physics.
CRC Press

\bibitem{ahn2018higher}
Ahn J, Yang BJ. 2019.
\textit{Phys. Rev. B} 99:235125

\bibitem{ahn2019stiefel}
Ahn J, Park S, Kim D, Kim Y, Yang BJ. 2019.
\textit{Chinese Physics B} 28:117101

\bibitem{bell2006gauss}
Bell D. 2006.
\textit{Journal of Geometry} 85:15--21

\bibitem{tqcsite}
 2019.
Topological materials database.
\url{https://topologicalquantumchemistry.org}

\bibitem{he2019symtopo}
He Y, Jiang Y, Zhang T, Huang H, Fang C, Jin Z. 2019.
\textit{Chinese Physics B} 28:087102

\bibitem{peri2020experimental}
Peri V, Song ZD, Serra-Garcia M, Engeler P, Queiroz R, et~al. 2020.
\textit{Science} 367:797--800

\bibitem{de2019engineering}
de~Paz MB, Vergniory MG, Bercioux D, Garc\'{\i}a-Etxarri A, Bradlyn B. 2019.
\textit{Phys. Rev. Research} 1:032005

\bibitem{alexandradinata2019crystallographic}
Alexandradinata A, H{\"o}ller J, Wang C, Cheng H, Lu L. 2019.
\textit{arXiv preprint arXiv:1908.08541}

\bibitem{watanabe2018structure}
Watanabe H, Po HC, Vishwanath A. 2018.
\textit{Science advances} 4:eaat8685

\bibitem{xu2020high}
Xu Y, Elcoro L, Song Z, Wieder BJ, Vergniory M, et~al. 2020.
\textit{arXiv preprint arXiv:2003.00012}

\bibitem{gooth2019evidence}
Gooth J, Bradlyn B, Honnali S, Schindler C, Kumar N, et~al. 2019.
\textit{Nature} 575:315--319

\bibitem{shi2019charge}
Shi W, Wieder BJ, Meyerheim H, Sun Y, Zhang Y, et~al. 2019.
\textit{arXiv preprint arXiv:1909.04037}

\bibitem{dmft}
Di~Sante D, Hausoel A, Barone P, Tomczak JM, Sangiovanni G, Thomale R. 2017.
\textit{Phys. Rev. B} 96:121106

\bibitem{rachel2018interacting}
Rachel S. 2018.
\textit{Reports on Progress in Physics} 81:116501

\bibitem{dzero2016topological}
Dzero M, Xia J, Galitski V, Coleman P. 2016.
\textit{Annual Review of Condensed Matter Physics} 7:249--280

\bibitem{khalaf2018higher}
Khalaf E. 2018.
\textit{Physical Review B} 97:205136

\bibitem{langbehn2017reflection}
Langbehn J, Peng Y, Trifunovic L, von Oppen F, Brouwer PW. 2017.
\textit{Physical review letters} 119:246401

\bibitem{wang2018weak}
Wang Y, Lin M, Hughes TL. 2018.
\textit{Physical Review B} 98:165144

\bibitem{bultinck2019three}
Bultinck N, Bernevig BA, Zaletel MP. 2019.
\textit{Physical Review B} 99:125149

\bibitem{ono2019symmetry}
Ono S, Yanase Y, Watanabe H. 2019.
\textit{Physical Review Research} 1:013012

\bibitem{schindler2020pairing}
Schindler F, Bradlyn B, Fischer MH, Neupert T. 2020.
\textit{arXiv preprint arXiv:2001.02682}

\end{thebibliography}
\end{document}